# PARAMETRIC INSTABILITIES AND PARTICLES HEATING OF CIRCULARLY POLARIZED ALFVÉN WAVES WITH AN INCOHERENT SPECTRUM: TWO-DIMENSIONAL HYBRID SIMULATIONS

Peng He[1]

[1] School of Earth and Space Sciences (SESS), University of Science and Technology of China (USTC), Hefei 230026, China

Corresponding Author: Peng He

Email: hepeng11@mail.ustc.edu.cn

# ABSTRACT

Plasma ions heating (especially minor heavy ions preferential heating) in fast solar wind and solar corona is an open question in space physics. However, Alfvén waves have been always considered as a candidate of energy source for corona heating. In this paper, by using a two-dimensional (2-D) hybrid simulation model in a low beta electron-proton-alpha plasma system, we have investigated the relationships between plasma ions heating and power spectra evolution of density and magnetic field fluctuations excited from the parametric instabilities of initial pump Alfvén waves with an incoherent spectrum at different propagation angles $\theta_{k_0 B_0}$ (an oblique angle between the initial pump wave vector $\mathbf{k}_0$ and the background magnetic field $\mathbf{B}_0$). It is found that, the wave-wave coupling as well as wave-particle interaction play key roles in ions heating, and an Alfvén spectrum with small propagation angle (e.g. $\theta_{k_0 B_0} = 15°$) can most effectively heat alpha particles in perpendicular direction as well as in parallel direction for both proton and alpha particle than the case of a monochromatic Alfvén wave or an Alfvén spectrum with larger propagation angle. Detailedly speaking, for the cases of monochromatic pump Alfvén wave, (1) the parallel temperature for each ion specie is enhanced as $\theta_{k_0 B_0}$ increasing to $\theta_{k_0 B_0} = 45°$ in early phase of the evolution while in latter phase almost decreases as $\theta_{k_0 B_0}$ increasing, (2) the growth profiles of the perpendicular temperature for alpha particles are roughly consistent with its parallel temperatures in early phase but the cases with smaller propagation angles $\theta_{k_0 B_0} = 0°, 15°$ exhibits more effectively increasing in latter phase (after about $\Omega_p t \approx 400$), while protons remains approximately constant on perpendicular temperature for all propagation angles. For the cases of pump Alfvén waves with an incoherent spectrum, (3) the ions parallel heating exhibits similar behavior to cases of monochromatic Alfvén wave and is enhanced more effectively to higher temperatures at the end of evolution especially for alpha particles, (4) the perpendicular temperature of alpha particles tends to increasing higher for cases with smaller propagation angles (especially for $\theta_{k_0 B_0} = 15°$) and is also significantly enhanced to higher values than the previous monochromatic case, still protons can be seldom heated in perpendicular direction. Moreover, for all above 2-D simulation cases, the ion (proton or alpha particle) beam is formed just parallel to the background magnetic field ($\mathbf{B}_0$) regardless of initial propagation angle of pump waves, which is mainly caused by Landau resonance between the ions and the excited ion-acoustic waves (IAWs) from the parametric instabilities in latter phase of the evolution, while the perpendicular heating of alpha particles mainly results from cyclotron resonance between the alpha particles and the high-frequency (large wavenumber) Alfvén waves (AWs) generated via wave-wave interactions in early or latter phase of the evolution. Finally, the observation of ion cyclotron waves (ICWs) with a wide range of wavenumbers and frequencies propagating at small angles with respect to the direction of the background magnetic field is discussed.

# 1. INTRODUCTION

The temperature of solar corona (about $10^6$ K) is almost 3 orders of magnitude higher than that of visible photosphere (Kohl et al. 2006), yet the energy source and heating mechanism of solar corona and solar wind have not been totally solved. However, in the chromosphere and solar wind large-amplitude Alfvén waves are observed ubiquitously (Belcher & Davis 1971; Belcher, Davis, & Smith 1969; Coleman 1966; De Pontieu et al. 2007; Roberts, Goldstein, & Klein 1990; Unti & Neugebauer 1968) and supported as a promising candidate for corona heating (Hanson & Voss 2007), which is often associated with the features of velocity distribution functions (VDFs) of solar wind ions. It is exhibited that, the protons are made up of a core component with a temperature anisotropy $T_{\perp p}/T_{\parallel p} > 1$ ( $\perp$ and $\parallel$ denote the directions perpendicular and parallel to the background magnetic field $\mathbf{B}_0$ ) and a field-aligned beamlike component with drift velocity $V_D \sim V_A$ (Feldman et al. 1973; Goldstein et al. 2000; Goodrich & Lazarus 1976; Marsch et al. 1982b; Tu, Marsch, & Qin 2004), while the alpha particles (~4% of the total particle number density) are streaming faster and hotter than the protons with $T_a \geq A_a T_p$ ( $A_a$ is the atom mass number of alpha particle) and a strong temperature anisotropy $T_{\perp a}/T_{\parallel a} \gg 1$ (Bourouaine et al. 2013; Isenberg & Hollweg 1983; Marsch 2006; Marsch et al. 1982a; Neugebauer et al. 1994; Reisenfeld et al. 2001; Von steiger et al. 1995). Although a circularly polarized Alfvén wave is an exact solution for both ideal incompressible magnatohydrodynamic (MHD) equations (Barnes & Hollweg 1974; Ferraro 1955) and full Vlasov-Maxwell equations with multi-species ions (Sonnerup

& Su 1967), it is still unstable to parametric decay instability with MHD description in a low beta plasma (Galeev & Oraevskii 1963; Sagdeev & Galeev 1969), which leads to energy transformations from the pump Alfvén wave ($k_0$) gradually to the decayed backward (opposite to $\mathbf{B}_0$) daughter Alfvén wave ($k^- = k_0 - k_s$, $k_0 < k_s$) and the forward ion acoustic wave ($k_s$) (Del Zanna, Velli, & Londrillo 2001; Jayanti & Hollweg 1993; Terasawa et al. 1986). Due to the dispersion effects of Hall-MHD, modulational instability can occur generating one forward ion acoustic wave ($k_s$) and two forward daughter Alfvén waves ($k^\pm = k_0 \pm k_s$, $k_0 > k_s$) (Hollweg 1994; Longtin & Sonnerup 1986; Sakai & Sonnerup 1983; Wong & Goldstein 1986). Further considering ion kinetic effects (finite ion temperature), compared with fluid description it can reduce the instability growth rates and broaden the unstable wave number range even in a low-beta plasma (Araneda, Marsch, & Vinas 2007; Inhester 1990; Kauffmann & Araneda 2008; Nariyuki & Hada 2006, 2007; Vasquez 1995). Especially, minor heavy ions (alpha particles) can make the plasma more sensible to kinetic effects (a small increment in the thermal energy of the alpha particles, in an already cold plasma, leads to a significant reduction of growth rate of decay instability) and can result in some new types of wave modes and instabilities (Gomberoff & Elgueta 1991; JOSEPH V. HOLLWEG 1993; Kauffmann & Araneda 2008).

As a feedback of ion kinetic effects, parametric instabilities can also affect ion dynamics in plasma system. Using a one-dimensional (1-D) hybrid simulation model, Araneda et al. (2008, 2009) have declared that a field-aligned proton beam can be produced via ion trapping with the ion acoustic waves (IAWs) excited from the modulational instability of a monochromatic Alfvén wave and alpha particles can be preferentially heated in perpendicular direction via pitch-angle-scattering transferring the initial fluid-motion energy to random ion-motion energy (Araneda, Maneva, & Marsch 2009; Araneda, Marsch, & A 2008), which provide a reasonable means for heating solar corona via nonlinear wave-wave interactions and wave-particle interactions starting from the parametric instabilities of Alfvén waves.

Therefore, with 1-D hybrid simulations several works have been devoted to investigating the ion dynamics in parametric processes and Alfvén turbulence (Gao et al. 2014; He et al. 2016; Maneva, Vinas, & Ofman 2013; Matteini et al. 2010b). Taking the initial pump Alfvén waves with an incoherent spectrum into account, Matteini (2010) have found that not only modulational but also decay instability can generate a field-aligned beam in proton VDFs via nonlinear ion trapping from the excited IAWs and He et al. (2016) have illustrated that alpha particles can be more effectively preferential heated in perpendicular direction than that of monochromatic case due to cyclotron resonance with the forward high-frequency Alfvén modes excited from parametric wave-wave coupling. Furtherly considering a solar wind expansion model and setting turbulent Alfvén-cyclotron spectra as initial pump waves, Maneva et al. (2013) have illustrated that, alpha particles can still be preferential heated by pitch-angle-scattering with the initial pump Alfvén-cyclotron spectra and differentially accelerated in latter phase of the evolution by proton beam pulling, in which the effects of the shape, amplitude, and type of initial wave-spectra are more important than that of solar wind expansion.

Besides the above 1-D hybrid simulation works, Maneva (2015) have performed 2-D hybrid simulations still with a solar wind expansion model to investigate alpha particles preferential heating by the oblique low frequency full developed Alfvén turbulent spectrum close to the observation data (Maneva et al. 2015). Then it is found that, alpha particles can be more effectively heated in both perpendicular and parallel directions than protons mainly due to the nonresonant particle scattering with low frequency transvers fluctuations from initial turbulent spectrum, especially the initial Alfvén turbulent spectrum with highly oblique propagation can more prominently provide the parallel electric field leading parallel heating of alpha particles.

However, in present paper, in order to illustrate the relationships between plasma ions heating and power spectra evolution of density and magnetic field fluctuations excited from the parametric instabilities of initial pump Alfvén waves with an incoherent spectrum at different propagation angles $\theta_{k_0 B_0}$, with 2-D hybrid simulations in a low beta electron-proton-alpha plasma system we focus on the nonlinear wave-wave and wave-particle interactions in the parametric processes and the results of ions heating without regard to the effects of solar wind expansion and drift velocity between ion species.

This paper is organized in four sections. In the following simulation model and simulation results are demonstrated in Section 2 and 3, respectively. Finally conclusions and discussion will be displayed in Section 4.

## 2. SIMULATION MODEL

A two-dimensional (2-D) hybrid simulation model with periodic boundary conditions is used to investigate plasma ions heating during the parametric instabilities of parallel or oblique propagating Alfvén waves with an incoherent spectrum in a low beta proton-electron-alpha plasma system, where the ions (protons and alpha particles) are described as particles and the electrons are treated as massless fluid (Quest 1988; Winske 1985; Winske & Omidi 1993). The simulations are performed in the $x-y$ plane, in which all the wave modes of the simulations propagate and the background magnetic field ($\mathbf{B}_0$) crosses with $x$ axis at an oblique angle $\theta_{k_0 B_0}$ expressed as $\mathbf{B}_0 = B_0(\cos\theta_{k_0 B_0}\hat{\mathbf{x}} + \sin\theta_{k_0 B_0}\hat{\mathbf{y}})$. In present simulations here, all the physical quantities depend on two spatial coordinate $x$ and $y$, while the velocity and electromagnetic field still have three components on $x$, $y$ and $z$ axes. For the convenience of numerical computation, the plasma density, magnetic field, and velocity are normalized by the initial uniform density $\rho_0$, background magnetic field $\mathbf{B}_0$, and background Alfvén velocity $V_A = B_0/\sqrt{\mu_0 \rho_0}$, respectively. Further the time and space are normalized by the reciprocal of proton cyclotron frequency $\Omega_p^{-1} = m_p/e_p B_0$ ($m_p$ is the mass of proton, and $e_p$ is the charge of proton) and the proton inertial length $c/\omega_{pp}$ ($c$ is the speed of light in a vacuum, and $\omega_{pp}$ is the proton plasma frequency based on the proton number density $n_p$), respectively.

The initial pump Alfvén waves with an incoherent spectrum propagating parallel to $x$ axis (equivalent to propagating at an an oblique angle $\theta_{k_0 B_0}$ with the

background magnetic field $\mathbf{B}_0$) are given as (He, et al. 2016; Matteini, et al. 2010b)

$$\delta \mathbf{B}_p = \sum_{k_0=k_1}^{k_n} \delta B_{k_0}[\cos(k_0 x - \omega_0 t + \varphi_{k_0})\hat{\mathbf{y}} + \sin(k_0 x - \omega_0 t + \varphi_{k_0})\hat{\mathbf{z}}] \quad (1)$$

$$\delta \mathbf{u}_i = \sum_{k_0=k_1}^{k_n} \delta u_{ik_0}[\cos(k_0 x - \omega_0 t + \varphi_{k_0})\hat{\mathbf{y}} + \sin(k_0 x - \omega_0 t + \varphi_{k_0})\hat{\mathbf{z}}] \quad (2)$$

, where $\delta \mathbf{B}_p$ is the transverse magnetic field fluctuation imposed by the pump Alfvén waves, and $\delta \mathbf{u}_i$ is the associated transverse velocity of the ions ($i = p, \alpha$, while $p$, $\alpha$ denotes proton and alpha particle, respectively). The initial phase $\varphi_{k_0}$ is given randomly in the range $[0, 2\pi]$. For each component of the initial pump Alfvén waves, the dispersion relation is given as $\omega_0 = k_0 \cos\theta_{k_0 B_0}$ (Gao et al. 2013; Matteini et al. 2010a), where $\omega_0$, $k_0$ represent the frequency and wave number of each initial Alfvén mode. According to the expressions in Eqs. (1) and (2), it is indicated that the initial pump Alfvén wave trains are a linear superposition of left-handed Alfvén waves with different wave numbers $k_0$ and propagate at an obliquely angle $\theta_{k_0 B_0}$. Moreover, $\delta B_{k_0}$ and $\delta u_{ik_0}$ satisfy the Walen's relation (Araneda, et al. 2009; Gao, et al. 2013; He, et al. 2016; Maneva, et al. 2015; Matteini, et al. 2010b):

$$\delta u_{ik_0} = -\frac{\omega_0 / k_0}{1 - \omega_0 / \Omega_i} \delta B_{k_0} \quad (3)$$

, where the initial transverse bulk velocity of alpha particles is larger than that of protons due to the different cyclotron frequencies.

At the beginning of the simulations, the ions (proton and alpha particle) satisfy the Maxwellian velocity distributions and the thermal velocity of alpha particles is the same as that of protons. For all simulation Runs here, the number of grid cells is $n_x = n_y = 150$, the size of grid cell is $\Delta x = 1.0 \, c/\omega_{pp}$, the time step is $\Delta t = 0.025 \, \Omega_p^{-1}$, and the electron resistive length is set to be $L_r = \eta/\mu_0 V_A = 0.02 \, c/\omega_{pp}$, which is much smaller than the grid size. Furthermore, 100 macroparticles of each ion specie are initially evenly distributed in every grid cell (for saving simulation time). The number density of alpha particles is set to be $n_a/n_e = 0.04$ (where $n_e = n_p + 2n_a$ is the number density of electrons). The electron beta is $\beta_e = 0.1$, while the proton beta is $\beta_p = 0.01$. The total magnetic field energy is set as $\sum_{k_0=k_1}^{k_n} (\delta B_{k_0}/B_0)^2 = 0.04$, and the amplitude $\delta B_{k_0}$ of each component for the initial pump Alfvén wave trains is equal. The wave number $k_0$ of each initial Alfvén mode is calculated as $k_0 = \dfrac{2\pi m_0}{n_x \cdot \Delta x}$ (where $m_0$ is the mode number for wave number $k_0$), then the corresponding frequency $\omega_0$ can be derived from Eq. (3).

## 3. SIMULATION RESULTS

For monochromatic cases, simulation Runs 1-5 have been performed with the pump Alfvén wave $k_0 = 0.209$ ($m_0 = 5$) and different oblique propagation angles $\theta_{k_0 B_0} = 0°, 15°, 30°, 45°, 60°$. While for spectrum cases, the pump Alfvén waves with an incoherent spectrum of $k_0 = 0.126, 0.168, 0.209, 0.251, 0.293$ ($m_0 = 3, 4, 5, 6, 7$) are initialized with different oblique propagation angles $\theta_{k_0 B_0} = 0°, 15°, 30°$ in simulation Runs 6-8. Initially in all above cases, the drift velocity between protons and alpha particles is $U_{\alpha p} = 0$.

First of all, we will present the simulation results of monochromatic cases with initial propagation angles $\theta_{k_0 B_0} = 0°$, $30°$ in Run 1 and Run 3 comparing the ions heating efficiency with the results of 1-D hybrid simulations (He, et al. 2016). Figure 1 displays the temporal evolutions of (a) the density fluctuations $<(\delta\rho/\rho_0)^2>^{1/2}$ and the temperatures of (b) protons and (c) alpha particles in Run 1 involving a monochromatic pump Alfvén wave $k_0 = 0.209$ ($m_0 = 5$) propagating parallel to the background magnetic field $\mathbf{B}_0$ ($\theta_{k_0 B_0} = 0°$). In panel (a) of Figure 1, it is clear that the density fluctuations experience a linear growth phase from about $\Omega_p t \approx 250$, which is proved to be a decay instability by power spectra analysis in Figure 2. Looking at panel (b) and (c) of Figure 1, it is shown that the parallel temperature for each ion specie exhibits an increasing phase in the evolution corresponding to the linear growth phase of the density fluctuations in panel (a) of Figure 1. However, the protons maintain almost constant temperatures in perpendicular direction while the alpha particles can be more effectively perpendicularly heated reaching to about 3.6 times of the initial temperature at $\Omega_p t = 2000$ than that of 1-D case (about 2.3 times of the initial temperature at $\Omega_p t = 2000$, not shown here). Here, the temperatures are calculated by the following two steps: in the first step, calculate the parallel temperature $T_{\|i} = m_i \langle (v_x - \langle v_x \rangle)^2 \rangle$ and the perpendicular temperature $T_{\perp i} = \frac{m_i}{2} \langle (v_y - \langle v_y \rangle)^2 + (v_z - \langle v_z \rangle)^2 \rangle$ of each grid cell (where the bracket $\langle \bullet \rangle$ denotes an average over all ions in one fixed grid cell for each ion specie); in the second step, average the above temperatures of one fixed grid over all grid cells for each ion specie. Such above algorithm can eliminate the effect of the bulk velocity in each grid cell (He, et al. 2016; Lu & Chen 2009), which differs from the 'apparent temperature' with the imposed bulk velocity from the magnetic modes (Araneda, et al. 2009; Maneva, et al. 2015; Matteini, et al. 2010b). In Figure 2, the power spectra in $k_x - k_y$ space for the density and magnetic field fluctuations is given at $\Omega_p t = 100, 470, 1000, 1800$ to analyze nonlinear wave-wave interactions of the parametric processes in Run 1. In the beginning (at about $\Omega_p t = 100$) of the evolution, because of a constant magnetic field intensity for monochromatic pump Alfvén wave in Run 1, there is no excited density fluctuations in panel (a) of Figure 2 with an initialized pump Alfvén mode $k_0 = 0.209$ propagating along $x$ axis (parallel to $\mathbf{B}_0$) in panel (b) of Figure 2. Consequently the linear growth phase of parametric decay appears at about $\Omega_p t = 470$. Along initial propagation direction (parallel to $\mathbf{B}_0$), pump Alfvén mode $k_0 = 0.209$ decays into a ion acoustic wave (IAW) $k_{s\|} \approx 0.34$ with several harmonic modes in panel (c) of Figure 2 and a daughter Alfvén wave (AW) $k_{\|}^- \approx -0.13$ cascading into high-frequency Alfvén modes around $k_{\|} \approx 0.5$ by wave-wave coupling between $k_0$ and $k_{s\|}$ in panel (d) of Figure 2, meanwhile the obvious transverse modulations occur in the direction perpendicular to $\mathbf{B}_0$ (Matteini, et al. 2010a). Comparing Figure 1 and Figure 2, it is clearly concluded that, the parallel heating of ions is caused by ions trapping via Landau resonance with the excited IAWs in the linear growth phase of parametric decay (Araneda, et al. 2009), but only minor heavy ions (alpha particle with lower gyrofrequency) can be effectively perpendicularly heated via cyclotron resonance with the cascaded high-frequency AWs around $k_\parallel \approx 0.5$. To directly show the behavior of ion dynamics, Figure 3 have plotted the scattering graph of velocity distribution in the ($v_x$, $v_y$) space at $\Omega_p t = 100, 470, 1000, 1800$, upper row for protons, bottom row for alpha particles. The proton beam start to generate in the linear growth phase of parametric decay at about $\Omega_p t = 470$ and keep it to the end of the evolution, while alpha particles begin perpendicular heating also corresponding to the linear growth phase of parametric decay with its beam and parallel heating produced latter after proton beam formation. Because of its heavy mass, only small number of alpha particles can be trapped by the IAWs difficultly leading remarkable alpha beams but still can be scattering heated in parallel direction at $\Omega_p t = 1000, 1800$ in panel (g) and (h) of Figure 3.

Considering oblique propagation case in Run 3 with the same parameters as Run1 but $\theta_{k_0 B_0} = 30°$, Figure 4 shows that the density fluctuations immediately increase with oscillations caused by the inhomogeneous magnetic field from oblique propagating initial pump Alfvén wave in panel (a) of Figure 4 and the parallel heating for each ion specie experiences two phases corresponding to different generating mechanics of IAWs in panel (b) and (c) of Figure 4. While the perpendicular heating of alpha particles occurs from the very beginning of the evolution, which is different from that of Run1. Yet in panel (b) of Figure 4, protons have no perpendicular heating as that of Run 1. Looking at Figure 5, in the very beginning at $\Omega_p t = 50$ IAW with its harmonic modes are immediately excited and meanwhile high-frequency AWs appear cascaded from wave-wave coupling along the propagation direction of initial pump Alfvén wave on $x$ axis (He, et al. 2016), then at $\Omega_p t = 400$ parametric decay instability occurs mainly along the propagation direction of initial pump Alfvén wave on $x$ axis immediately with perpendicular decay and transverse modulation along the direction perpendicular to $\mathbf{B}_0$ (Gao, et al. 2013; Matteini, et al. 2010a), latterly at $\Omega_p t = 1000, 1500$ the density and magnetic field fluctuations are gradually dominated by the oblique modes with more transverse modulations and wave numbers shrinking toward small values. According to the results from Figure 4 and 5 for oblique propagating monochromatic pump Alfvén wave, we draw a conclusion that, the parallel heating for each ion specie is still mainly caused by Landau resonance with the excited IAWs but experiences two growth phases different from Run 1, while the perpendicular heating of alpha particles results from cyclotron resonance with the high-frequency AWs cascaded from wave-wave coupling between IAWs and AWs. Figure 6 supplements Figure 4 to illustrate the ions dynamic behavior, in which proton beam is produced along $\mathbf{B}_0$ (oblique to $x$ axis with about $\theta_{k_0 B_0} = 30°$) mainly due to the parallel component of the electric field of the excited IAWs from two different phases of the evolution as well as alpha beam latter generated, while the perpendicular heating of alpha particles increases as soon as the high-frequency Alfvén modes appear from the very beginning of the evolution.

Furthermore the oblique propagation effects are investigated in Run 1-5 with different oblique propagation angles $\theta_{k_0 B_0} = 0°, 15°, 30°, 45°, 60°$ giving the temporal evolutions of ion temperatures plotted together in Figure 7. Panel (a) and (c) show that, the parallel heating for each ion specie roughly experiences two phases, which is enhanced higher as $\theta_{k_0 B_0}$ increasing to $\theta_{k_0 B_0} = 45°$ in early phase of the evolution (before about $\Omega_p t \approx 400$) and recedes with $\theta_{k_0 B_0}$ increasing in latter phase. From the view of the above plots, it is inferred that the parallel heating of ions is dominated by the parallel projection of the electric field caused by the IAWs mainly on $x$ axis, which are excited by the inhomogeneous magnetic field of oblique propagating pump Alfvén wave in early phase and the parametric decay instability in latter phase, respectively. Then panel (d) tells us that the perpendicular heating of alpha particles also presents two growth phases, in which alpha particles are heated mainly via the perpendicular projection of the transverse electric field of the cascaded high-frequency AWs. For early phase (before about $\Omega_p t \approx 400$), the intensity of the cascaded AWs are chiefly determined by the couplings between initial Alfvén mode and the excited IAWs, which leads the growth profiles of perpendicular temperatures of alpha particles (in panel (d) of Figure 7) are consistent with those of parallel temperatures (in panel (c) of Figure 7). While for latter phase in panel (d) of Figure 7, the excited parametric decay and transverse modulation can modify the intensity of the cascaded AWs disordering the growth profiles of perpendicular temperatures compared with early phase, which indicates heating mechanism competitions between early and latter phases. Detailedly speaking in panel (d) of Figure 7, the perpendicular heating of alpha particles in the cases with smaller propagation angles ($\theta_{k_0 B_0} = 0°$, $15°$) is dominated by the parametric processes in latter phase, while in cases with larger propagation angles ($\theta_{k_0 B_0} = 30°$, $45°$) that is controlled by the intensity of the excited

IAWs (leading the cascaded AWs by wave-wave coupling) in early phase. However in the case with largest propagation angle ($\theta_{k_0 B_0} = 60°$), the perpendicular heating of alpha particles is suppressed in both early and latter phases. Due to its larger gyrofrequency, proton can hardly resonate with the cascaded AWs and maintains its perpendicular temperature almost as constant in panel (b) of Figure 7.

Secondly, the pump Alfvén waves with an incoherent spectrum ($k_0 = 0.126, 0.168, 0.209, 0.251, 0.293$) have been initialized in Run 6-8 as well as considering different propagation angles $\theta_{k_0 B_0} = 0°, 15°, 30°$, respectively. For the case of parallel propagating Alfvén spectrum in Run 6, Figure 8 shows that, the density fluctuations increase immediately from the very beginning due to the ponderomotive force caused by the initial Alfvén spectrum (He, et al. 2016; He et al. 2015; Matteini, et al. 2010b; Nariyuki, Hada, & Tsubouchi 2007) in panel (a) and the parallel temperature for each ion specie starts to increase at about $\Omega_p t \approx 300$ in panel (b) and (c) corresponding to the linear growth phase of parametric decay instability in panel (c) and (d) of Figure 9. Still looking at Figure 8, it is revealed that, the perpendicular heating of alpha particles has been effectively enhanced to about 4.6 times of the initial perpendicular temperature at $\Omega_p t = 2000$ higher than that of Run 1 (about 3.6 times of the initial perpendicular temperature at $\Omega_p t = 2000$, shown in panel (c) of Figure 1), while protons also keep almost constant perpendicular temperature similar to Run 1. Further Figure 9 plots the power spectra of the density and magnetic field fluctuations in $k_x - k_y$ space at $\Omega_p t = 100, 600, 1000, 1600$ for Run 6. It is found that the density spectrum along $x$ axis is immediately produced from the very beginning at $\Omega_p t = 100$ due to the ponderomotive force from initial Alfvén spectrum in panel (a) of Figure 9 meanwhile the initial Alfvén spectrum cascades to higher frequency (larger wave vectors intensively along $x$ axis) vis wave-wave couplings in panel (b) of figure 9 (He, et al. 2016). After Alfvén spectrum cascading, parametric decay occurs still mainly on $x$ axis with evident transverse modulations (Matteini, et al. 2010a) at $\Omega_p t = 600$ in panel (c) and (d) of Figure 9. For latter phase of the evolution at $\Omega_p t = 1000, 1600$ in Figure 9, the density and magnetic field spectra gradually broaden to large wave number in perpendicular direction while shrink to small wave number in parallel direction. Moreover, Figure 10 directly shows the ions dynamics that, the proton and alpha particle beams are gradually excited as IAWs enhanced with parametric decay along $\mathbf{B}_0$, while in panel (g) and (h) of Figure 10 alpha particles are more scattering heated in perpendicular direction than that of monochromatic case in Run 1, which is because of the dispersion of the excited high-frequency Alfvén spectrum via the spectrum couplings (Lu & Wang 2006). Back considering Figure 8, although the density fluctuations have been excited at the very beginning in panel (a), the parallel temperatures of ions have no effective growth before $\Omega_p t \approx 300$ in panel (b) and (c), which can be interpreted by the dispersion relations of the density and magnetic field fluctuations in Figure 11 that in early phase IAWs are excited from the modulation instability caused by ponderomotive force with higher phase velocity $\sim 0.7 V_A$ and only can resonate with few ions (Araneda, et al. 2009).

Then taking the effects of oblique propagation into account in Run 8 with the same parameters as Run 6 but $\theta_{k_0 B_0} = 30°$, the temporal evolutions of density fluctuations and ion temperatures, the power spectra analysis in the $k_x - k_y$ space are plotted in Figure 12 and 13, respectively. It is illustrated that, the density fluctuations suddenly increase to the highest value then keep oscillating to the end of the evolution in panel (a) of Figure 12, the parallel heating for each ion specie is enhanced from the very beginning mainly due to the excited IAWs in early phase (before about $\Omega_p t \approx 200$) and then experiences two growth phases in panel (b) and (c) of Figure 12, while in panel (c) of Figure 12 the perpendicular heating for alpha particles is significant dominated before about $\Omega_p t \approx 800$ because of the intensive cascaded high-frequency AWs corresponding to Figure 13 and 15 (before about $\Omega_p t \approx 800$). Moreover, Figure 13 shows that the power spectra evolution of density and magnetic field fluctuations in $(k_x, k_y)$ space for Run 8 is roughly consistent with that of Run 6 in Figure 9. But for oblique propagation case in Run 8, the excited density fluctuations in early phase (before $\Omega_p t \approx 200$) in panel (a) of Figure 13 exist small wave number modes with lower phase velocity ($\sim 0.2 V_A$) corresponding to panel (a) of Figure 15, which can be resonant with a number of ions leading immediately growth of its parallel temperature (before $\Omega_p t \approx 200$), and the spectrum of perpendicular decay and transverse modulations along the direction perpendicular to $\mathbf{B}_0$ are also produced in latter phase of figure 13. Further in figure 14, proton and alpha beams are still evidently observed along $\mathbf{B}_0$ cross with $x$ axis at about $\theta_{k_0 B_0} = 30°$ and generated earlier than that of parallel propagation case in Run 6, while alpha particles are still scattering perpendicularly heated. According to the dispersion relations of density and magnetic field fluctuations in Figure 15, it is found that, in early phase (before about $\Omega_p t \approx 200$) the density modes with higher ($\sim 0.71 V_A$) and lower ($\sim 0.2 V_A$) phase velocities are both excited, while in latter phase (after about $\Omega_p t \approx 200$) parametric decay instabilities begin to occur because of the magnetic modes with negative frequency appearing in panel (d) of Figure 15. Parametric decay generates new density modes with lower ($\sim 0.2 V_A$) phase velocity and magnetic modes cascading to higher frequency still occurs in panel (c) and (d) of Figure 15.

At the end of present paper, to investigate the oblique propagation effects for the cases of pump Alfvén waves with an incoherent spectrum, Run 6-8 have been performed and Figure 16 has plotted the temporal evolutions of ion temperatures putting simulation results with different propagation angles $\theta_{k_0 B_0} = 0°, 15°, 30°$ together. It is found that the growth profiles of parallel and perpendicular temperatures for both protons and alpha particles are roughly similar to that of cases for monochromatic pump Alfvén wave in Run 1-5. In early phase (before about $\Omega_p t \approx 200$) the parallel heating for each ion specie increases higher as $\theta_{k_0 B_0}$ increasing to $\theta_{k_0 B_0} = 45°$, but in latter phase (after about $\Omega_p t \approx 1000$) the parallel heating of ions is more effective for cases with smaller propagation angles $\theta_{k_0 B_0} = 0°, 15°$ growing to higher temperatures. However for perpendicular heating of alpha particles, it is also effectively heated in cases with smaller propagation angles $\theta_{k_0 B_0} = 0°, 15°$ according to the total heating effects of both early and latter phases in the whole evolution. Further, comparing Figure 7 and Figure 16, it is indicated that, pump Alfvén waves with an incoherent spectrum can heat ions (proton and alpha particle) more effectively in either parallel or perpendicular direction (respect to $\mathbf{B}_0$) than monochromatic pump Alfvén wave. Furthermore, Alfvén spectrum with small propagation angle (e.g. $\theta_{k_0 B_0} = 15°$) can lead the most effectively ions heating in both parallel and perpendicular directions, which is caused by the combined effects of ion heating in early and latter phases.

## 4. CONCLUSIONS AND DISCUSSION

In this paper, a 2-D hybrid simulation model is employed in a low beta electron-proton-alpha plasma system to detailedly analyze the relationships between plasma ions heating and power spectra evolution of density and magnetic field fluctuations excited from the parametric instabilities of initial pump Alfvén waves with an incoherent spectrum at different propagation angles $\theta_{k_0 B_0}$, and wave-wave coupling as well as wave-particle interaction are proved to play key roles in ions heating. Here the processes of ions heating are illustrated as following steps. At first via wave-wave coupling or parametric processes, IAWs and resonant high-frequency AWs are generated. Then by Landau resonance with the excited IAWs, ions can be heated and accelerated in parallel direction, while by cyclotron resonance with the excited resonant high-frequency AWs, alpha particles can be perpendicularly heated. According to the above heating processes, for pump Alfvén waves with an incoherent spectrum or monochromatic wave in 2-D simulation space of this paper, the ions heating experiences two main temporal growth phases, in which for early phase

(before about $\Omega_p t \approx 200 \sim 400$) the excited IAWs caused by the envelop modulations from the inhomogeneous magnetic field dominate the ions parallel heating and for latter phase the parametric decay producing new IAWs can modify or control the growth profiles of parallel temperatures. However for perpendicular heating of alpha particles the intensities of wave-wave interactions between IAWs and IAWs determine the growth profiles of perpendicular temperatures.

In the plasma system, due to the imposed pump Alfvén spectrum with different propagation angles, the wave-wave interactions become more complex leading more combined processes of ions heating via wave-particle interactions. Therefore, we summarized the results of ions heating in this paper as follow. For the cases of monochromatic pump Alfvén wave, (1) the parallel temperature for each ion specie is enhanced as $\theta_{k_0 B_0}$ increasing to $\theta_{k_0 B_0} = 45°$ in early phase of the evolution while in latter phase almost decreases as $\theta_{k_0 B_0}$ increasing in panel (a) and (c) of Figure 7, (2) the growth profiles of the perpendicular temperature for alpha particles are roughly consistent with its parallel temperatures in early phase but the perpendicular temperature of alpha particles in cases with smaller propagation angles $\theta_{k_0 B_0} = 0°, 15°$ exhibits more effectively increasing in latter phase (after about $\Omega_p t \approx 400$) in panel (d) of Figure 7, while protons remains approximately constant on perpendicular temperature for all propagation angles in panel (b) of Figure 7. For the cases of pump Alfvén waves with an incoherent spectrum, (3) the parallel heating for each ion specie exhibits similar behavior to cases of monochromatic Alfvén wave but is enhanced more effectively to higher temperatures at the end of evolution especially for alpha particles in panel (c) of Figure 16, (4) the perpendicular temperature of alpha particles tends to increasing higher for cases with smaller propagation angles (especially for smaller oblique propagation angle $\theta_{k_0 B_0} = 15°$) in panel (d) of Figure 16 and is also significantly enhanced to higher values than the previous monochromatic case, still protons can be seldom heated in perpendicular direction in panel (b) of Figure 16. As a result of ions heating and acceleration, considering ion velocity distribution functions (VDFs) for all above cases in 2-D simulation space, it is found that, (1) the evident ion (proton or alpha particle) beams tend to generate intensively parallel to the background magnetic field $\mathbf{B}_0$ (even for cases of initial oblique propagating pump Alfvén waves) in latter phase of the evolution, which is mainly dominated by Landau resonance between the ions and the excited IAWs from parametric decay instabilities, (2) while the scattering of perpendicular velocities in alpha particle VDFs mainly results from cyclotron resonance between the alpha particles and the high-frequency (large wavenumber) AWs generated via wave-wave interactions in early or latter phase of the evolution, especially more high-frequency Alfvén modes can be excited in the cases of initial pump Alfvén waves with a spectrum structure leading more effectively perpendicular heating of alpha particles. From the view of the above conclusions, when 2-D simulation space is taken into account (Matteini, et al. 2010a) and Alfvén waves with an incoherent spectrum (He, et al. 2016) at different propagation angles are initially imposed in a low beta electron- proton-alpha plasma system, it is found that, an incoherent Alfvén spectrum with smaller oblique propagation angle (e.g. $\theta_{k_0 B_0} = 15°$) can most effectively heat alpha particles in perpendicular direction as well as in parallel direction (in panel (c) and (d) of Figure 16) than the case of a monochromatic Alfvén wave or an Alfvén spectrum with larger propagation angle.

By using a 1-D hybrid simulation model and pump Alfvén waves with an incoherent spectrum, He et al. (2016) have illustrated that the perpendicular heating of alpha particles experiences two effective heating phases via cyclotron resonance with cascaded high-frequency AWs (He, et al. 2016). However, when a 2-D simulation model introduced to present paper, in panel (d) of Figure 7 and Figure 16 the perpendicular heating of alpha particles in latter phase becomes less robust compared with 1-D case in He et al. (2016). Although parametric decay instabilities still occur in latter phase for present paper, the cascading between IAWs and AWs exhibits rather weaker leading perpendicular heating of alpha particles weaker. It is probably because of the parametric decay instabilities in latter phase transmitting pump energy to extra transvers modulation modes in 2-D simulations (Matteini, et al. 2010a).

Ions heating in present work is always considered to be the combined effects of heating processes in different phases of the evolution. Moreover, the heating processes exhibit competitions between different phases influencing the final effects of ion heating. In panel (c) of Figure 7, the parallel heating of alpha particles is controlled by the combined effects, which contain the parallel electric field from the excited IAWs in early phase (its intensity increasing as $\theta_{k_0 B_0}$ increasing to $\theta_{k_0 B_0} = 45°$) and in latter phase (its intensity decreasing as $\theta_{k_0 B_0}$ increasing) as well as the parallel projection of transverse electric field from cascaded AWs in latter phase (smaller $\theta_{k_0 B_0}$ leading robust AWs cascading and resonant transverse electric field). For the case with $\theta_{k_0 B_0} = 15°$ (red dot line in panel (c) of Figure 7), its combined effects of parallel heating happen to meet the largest value than other cases. In addition, for perpendicular heating of alpha particles in panel (d) of Figure 7 and Figure 16, the cases with smaller propagation angles (e.g. $\theta_{k_0 B_0} = 0°, 15°$) display more effectively heating due to more robust wave-wave interactions between IAWs and AWs than other cases with larger propagation angles. However, detailedly comparing the competition effects from different heating processes require more theoretical and quantitative investigations, which might be performed in our future research.

Recently, from solar wind data of STEREO and MESSENGER Jian et al. (2009, 2010) have found that, the parallel and oblique ion cyclotron waves (ICWs) exist in the space from 0.3 to 1 AU with a wide range of wavenumbers and frequencies, and a statistically significant fraction of events propagate at small angles with respect to the direction of the background magnetic field (Jian et al. 2010; Jian et al. 2009). Therefore, from the view of this paper, in the solar wind from 0.3 to 1 AU, the heating processes for an Alfvén spectrum (a little lower frequency than ICWs) with small propagation angles can interpret and dominate the perpendicular heating of alpha particles as well as parallel heating of ions.

# Acknowledgments

This work was supported by the National Science Foundation of China, Grant Nos. 41474125, 41331067, 11235009, 41527804, 41421063, 973 Program (2012CB825602, 2013CBA01503).

# Figure Captions

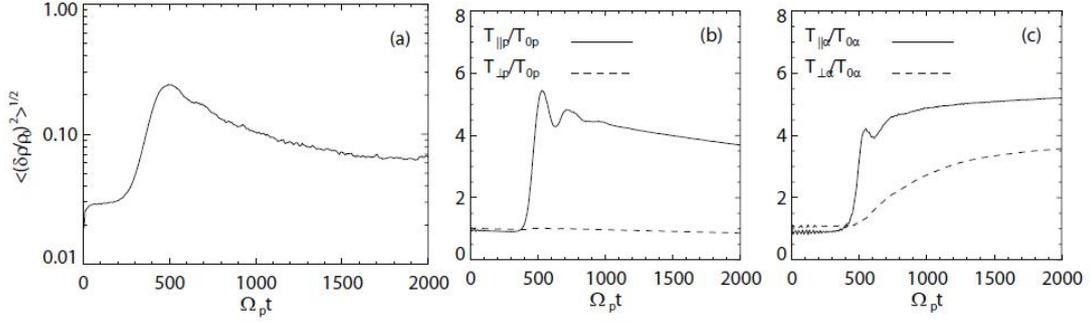

**Figure 1.** Time evolutions of (a) the density fluctuations $<(\delta\rho/\rho_0)^2>^{1/2}$ and the temperatures of (b) protons and (c) alpha particles (solid line for parallel temperature while dashed line for perpendicular temperature) in Run 1 with $\theta_{k_0 B_0} = 0°$.

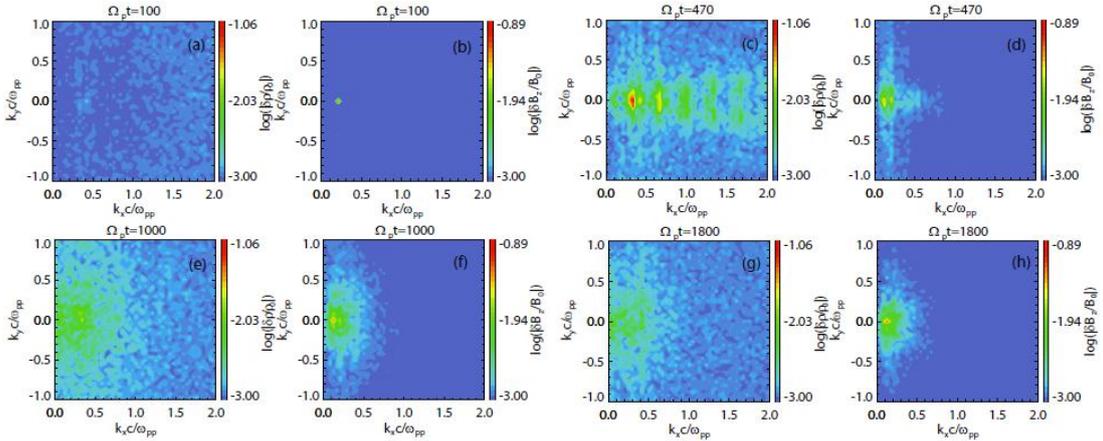

**Figure 2.** Power spectra of the density and magnetic field fluctuations at $\Omega_p t = 100, 470, 1000, 1800$ in the $(k_x, k_y)$ space by FFT transforming in Run 1 with $\theta_{k_0 B_0} = 0°$.

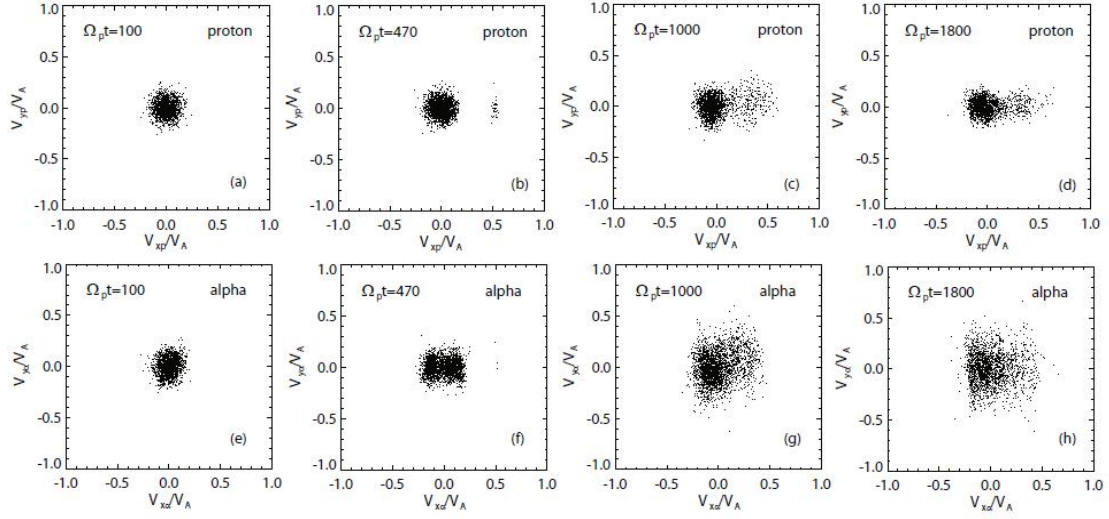

**Figure 3.** Scatter plots of ions velocities at $\Omega_p t = 100, 470, 1000, 1800$ in the ($v_x, v_y$) space, upper row for protons while bottom row for alpha particles in Run 1 with $\theta_{k_0 B_0} = 0°$.

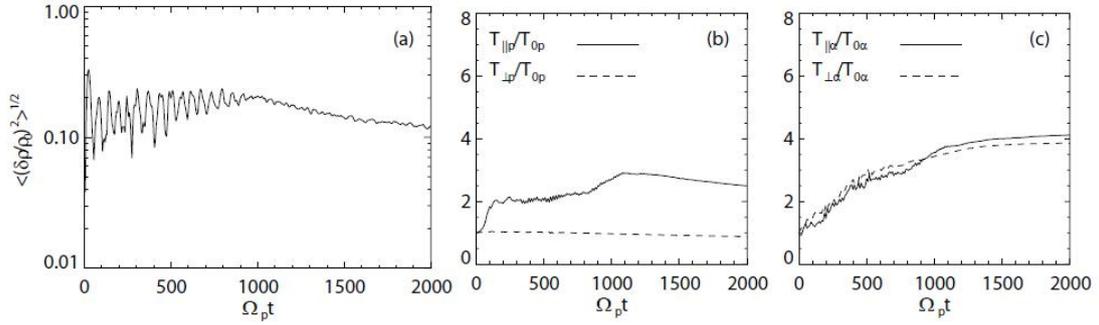

**Figure 4.** Time evolutions of (a) the density fluctuations $<(\delta\rho/\rho_0)^2>^{1/2}$ and the temperatures of (b) protons and (c) alpha particles (solid line for parallel temperature while dashed line for perpendicular temperature) in Run 3 with $\theta_{k_0 B_0} = 30°$.

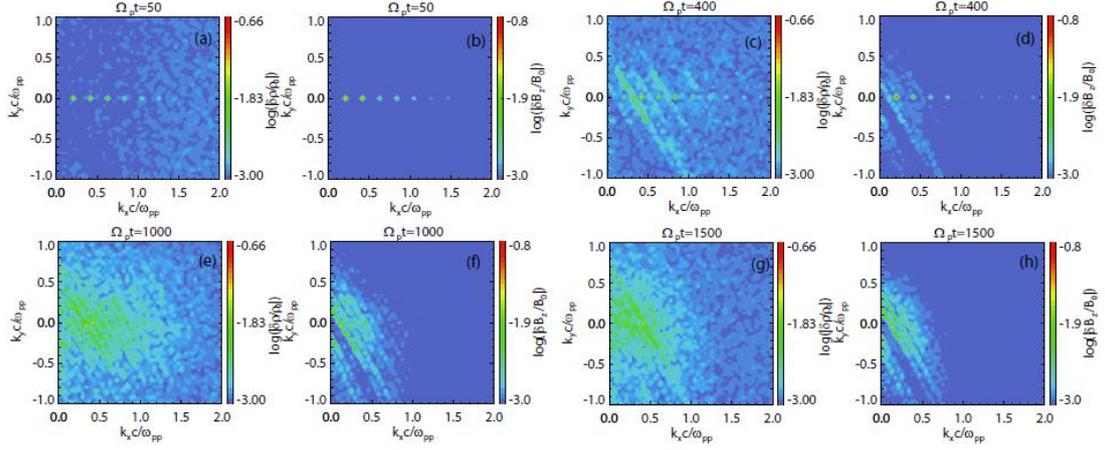

**Figure 5.** Power spectra of the density and magnetic field fluctuations at $\Omega_p t = 50, 400, 1000, 1500$ in the $(k_x, k_y)$ space by FFT transforming in Run 3 with $\theta_{k_0 B_0} = 30°$.

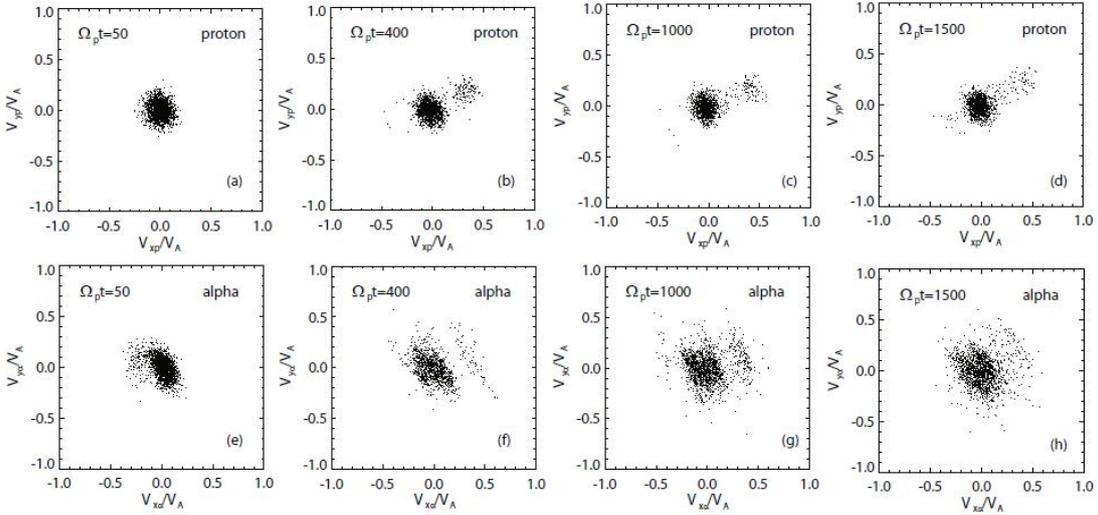

**Figure 6.** Scatter plots of ions velocities at $\Omega_p t = 50, 400, 1000, 1500$ in the $(v_x, v_y)$ space, upper row for protons while bottom row for alpha particles in Run 3 with $\theta_{k_0 B_0} = 30°$.

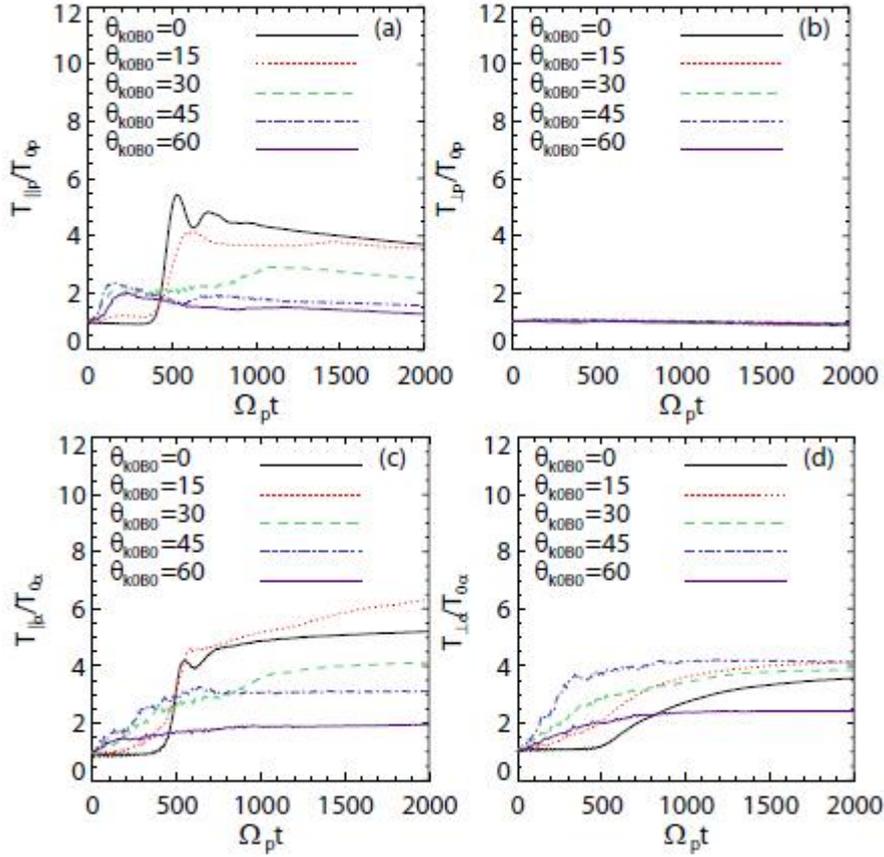

**Figure 7.** Time evolutions of the parallel and perpendicular temperatures in Run 1-5 with $\theta_{k_0 B_0} = 0°, 15°, 30°, 45°, 60°$ (denoted by black solid, red dot, green dashed, blue dashed dot and purple solid lines, respectively), upper row for protons while bottom row for alpha particles.

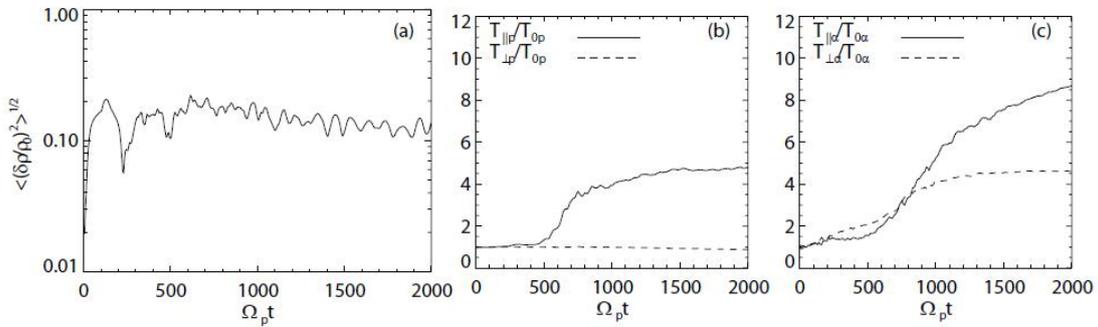

**Figure 8.** Time evolutions of (a) the density fluctuations $<(\delta\rho/\rho_0)^2>^{1/2}$ and the temperatures of (b) protons and (c) alpha particles (solid line for parallel temperature while dashed line for perpendicular temperature) in Run 6 with $\theta_{k_0 B_0} = 0°$.

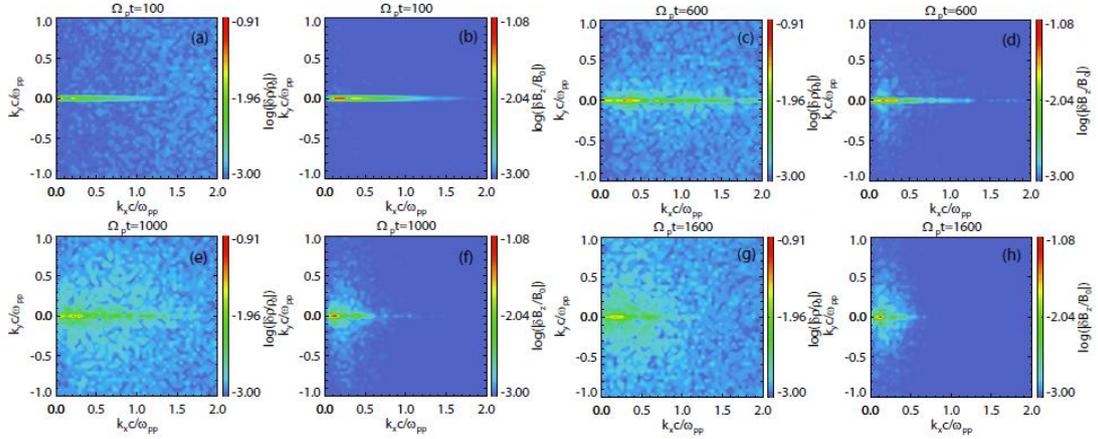

**Figure 9.** Power spectra of the density and magnetic field fluctuations at $\Omega_p t = 100, 600, 1000, 1600$ in the $(k_x, k_y)$ space by FFT transforming in Run 6 with $\theta_{k_0 B_0} = 0°$.

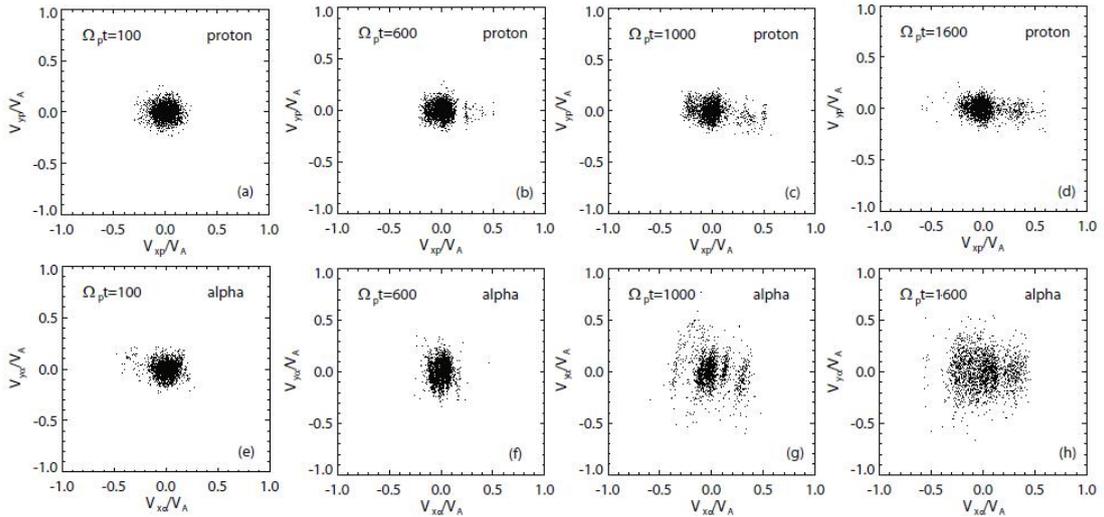

**Figure 10.** Scatter plots of ions velocities at $\Omega_p t = 100, 600, 1000, 1600$ in the $(v_x, v_y)$

space, upper row for protons while bottom row for alpha particles in Run 6 with $\theta_{k_0 B_0} = 0°$.

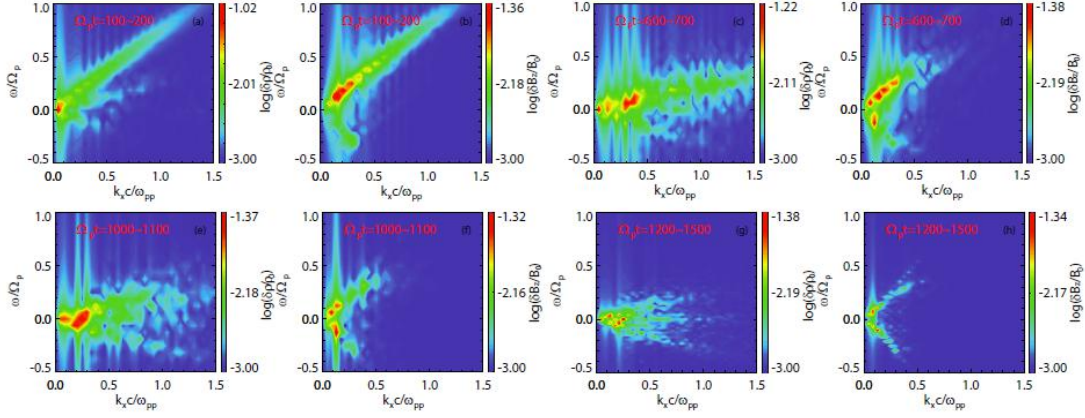

**Figure 11.** Dispersion relation of the density and magnetic field fluctuations on time intervals of $\Omega_p t = 100 \sim 200$, $600 \sim 700$, $1000 \sim 1100$, $1200 \sim 1500$ in the ($\omega$, $k_x$) space by FFT transforming in Run 6 with $\theta_{k_0 B_0} = 0°$.

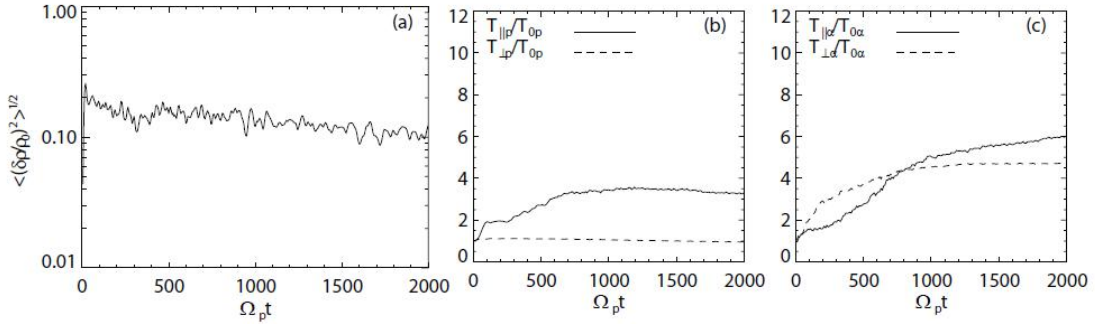

**Figure 12.** Time evolutions of (a) the density fluctuations $<(\delta\rho/\rho_0)^2>^{1/2}$ and the temperatures of (b) protons and (c) alpha particles (solid line for parallel temperature while dashed line for perpendicular temperature) in Run 8 with $\theta_{k_0 B_0} = 30°$.

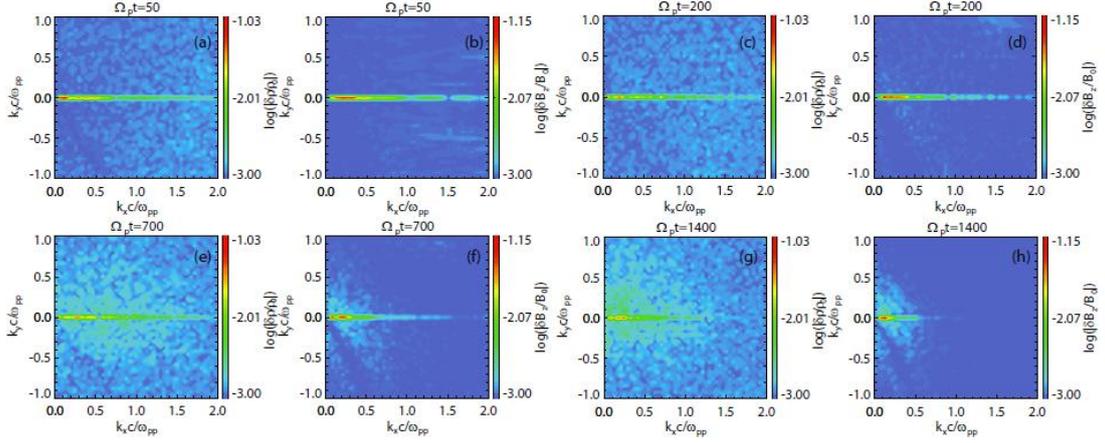

**Figure 13.** Power spectra of the density and magnetic field fluctuations at $\Omega_p t = 50, 200, 700, 1400$ in the $(k_x, k_y)$ space by FFT transforming in Run 8 with $\theta_{k_0 B_0} = 30°$.

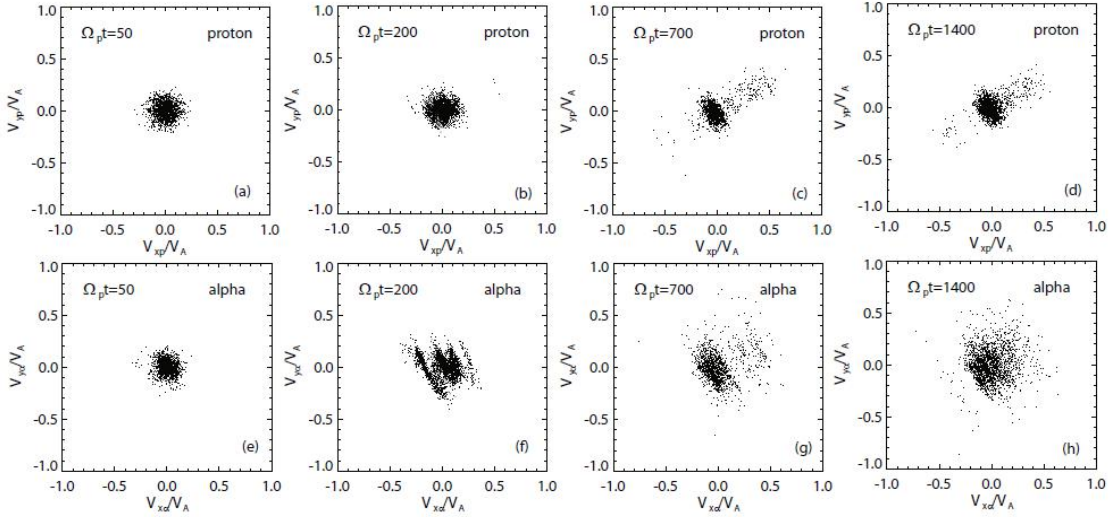

**Figure 14.** Scatter plots of ions velocities at $\Omega_p t = 50, 200, 700, 1400$ in the $(v_x, v_y)$ space, upper row for protons while bottom row for alpha particles in Run 8 with $\theta_{k_0 B_0} = 30°$.

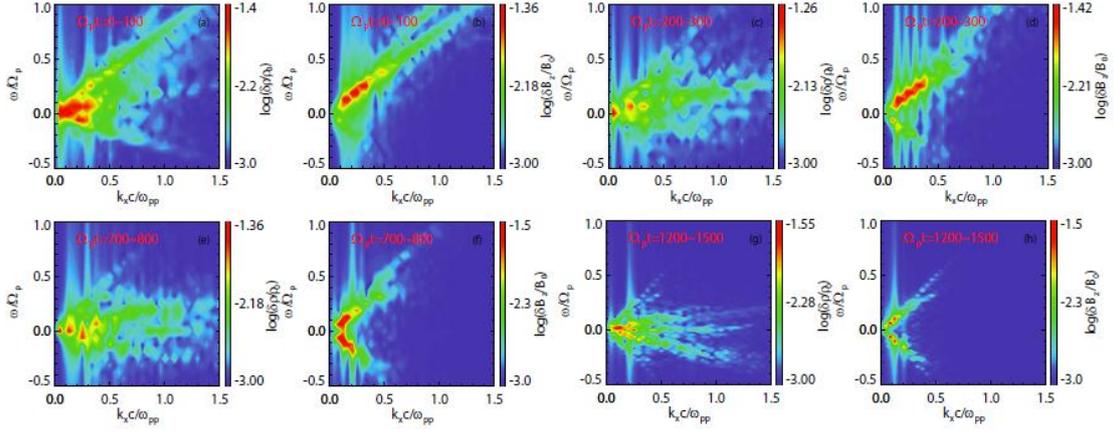

**Figure 15.** Dispersion relation of the density and magnetic field fluctuations on time intervals of $\Omega_p t = 0 \sim 100,\ 200 \sim 300,\ 700 \sim 800,\ 1200 \sim 1500$ in the ($\omega, k_x$) space by FFT transforming in Run 8 with $\theta_{k_0 B_0} = 30°$.

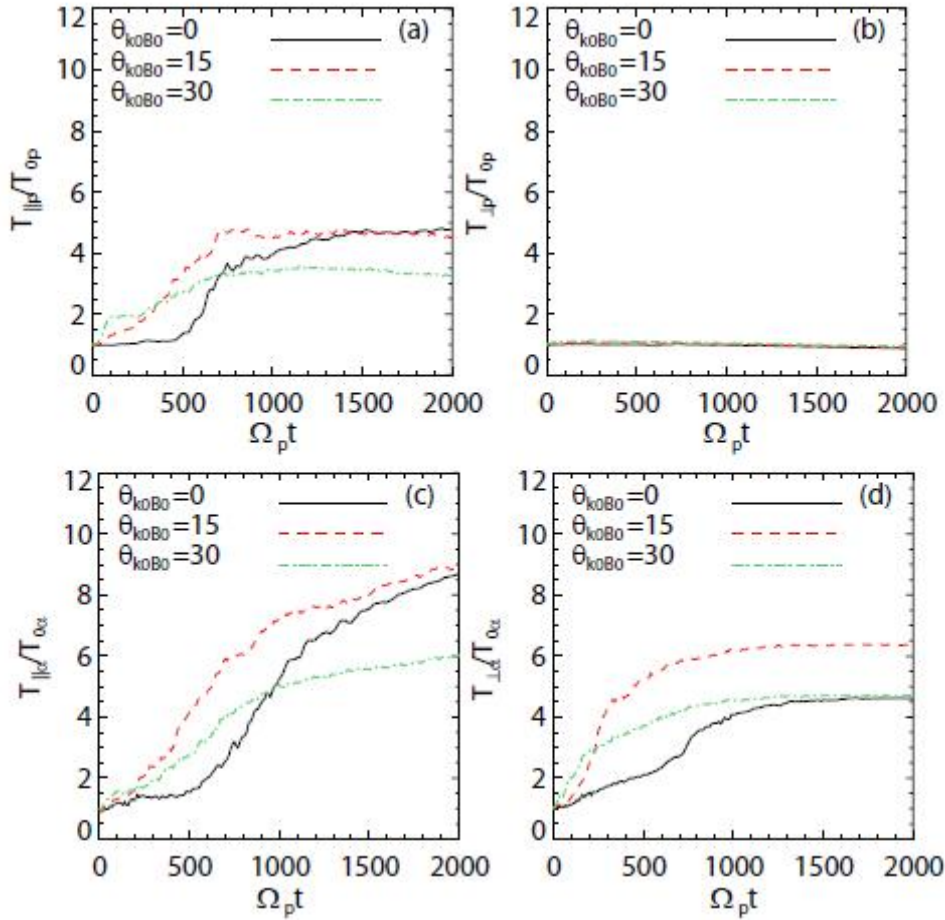

**Figure 16.** Time evolutions of the parallel and perpendicular temperatures in Run 6-8 with $\theta_{k_0 B_0} = 0°, 15°, 30°$ (denoted by black solid, red dashed and green dashed dot lines, respectively), upper row for protons while bottom row for alpha particles.